\documentclass[prl,twocolumn,aps]{revtex4-1}

\usepackage{amssymb}
\usepackage{graphicx}

\begin{document}
\title{Multi-band effects in in-plane resistivity anisotropy of\\ strain-detwinned disordered Ba(Fe$_{1-x}$Ru$_{x}$)$_{2}$As$_{2}$}

\author{E. C. Blomberg}
\author{M. A. Tanatar}
\author{A. Thaler}
\author{S. L. Bud'ko}
\author{P. C. Canfield}
\author{R. Prozorov}
\affiliation{Ames Laboratory and Department of Physics \& Astronomy, Iowa State University, Ames, Iowa 50010}

\date{Received: 3 April 2018; Published: 11 July 2018}

\begin{abstract}
In-plane resistivity anisotropy was measured in strain-detwinned as-grown and partially
annealed samples of isovalently-substituted $\mathrm{Ba(Fe_{1-x}Ru_{x})_{2}As_{2}}$ ($0<x \leq 0.125$) and the results were contrasted with previous reports on anneal samples with low residual resistivity. In samples with high residual resistivity, detwinned with application of strain, the difference of the two components of in-plane resistivity in the orthorhombic phase, $\rho_a -\rho_b$,  was found to obey Matthiessen rule irrespective of sample composition, which is in stark contrast with observations on annealed samples. Our findings are consistent with two-band transport model in which contribution from high mobility carriers of small pockets of the Fermi surface has negligible anisotropy of residual resistivity and is eliminated by disorder. Our finding suggests that magnetic/nematic order has dramatically different effect on different parts of the Fermi surface. It predominantly affects inelastic scattering for small pocket high mobility carriers and elastic impurity scattering for larger sheets of the Fermi surface.
\textbf{Published:} Journal of Physics: Condensed Matter \textbf{30} 315601 (2018).
\end{abstract}

\maketitle

Upon cooling, the parent compounds of iron-arsenide superconductors,
$\mathrm{AFe_{2}As_{2}}$ ($\mathrm{A=Ba,Ca,Sr}$), undergo a structural
phase change from a tetragonal to an orthorhombic crystal lattice
at a characteristic temperature $T_{s}$~\cite{Ca-phasetransition}.
This structural symmetry lowering is usually accompanied or followed by magnetic ordering below $T_m$ from a paramagnetic to a stripe type antiferromagnetic (AFM) state ~\cite{Cruz}. The trend for coupled but split orthorhombic/magnetic transitions is quite general for underdoped compounds of iron based materials, including NdFeAsO (1111 family), NaFeAs (111 family), with FeSe (11 family) being the only exception where orthorhombic distortion is followed by magnetic ordering only under pressure \cite{FeSeKothapati}.

In contrast to minute magnitude of the orthorhombic distortion (typically less than 0.5\% \cite{Blombergstrain}), large in-plane anisotropy of the normal-state electronic properties in the orthorhombic/antiferromagnetic phase of iron-based superconductors has been demonstrated by a variety of experimental probes, such as resistivity \cite{detwinning,FisherScience1}, thermopower \cite{thermopower,thermopower2}, Nernst effect \cite{thermopower2,Nernst}, optical conductivity \cite{optics1,optics2}, neutron scattering \cite{neutrons}, Raman scattering \cite{Raman}, scanning tunneling microscopy \cite{AllanSTM}, ARPES \cite{FisherReview},
and ultrasound \cite{ultrasound}, see Refs.~\onlinecite{FisherReview,RFMReview} for review.
It was argued that large anisotropy is driven by the electronic degree of freedom, referred to as electronic nematic instability \cite{RFMReview}. There is though intense debate as to the exact mechanism the electronic anisotropy appears.

The electrical conductivity of a metal, $\sigma=e(n/m^*)\tau $, depends both on parameters of the Fermi surface, reflected in the $n/m^*$ ratio known as the Drude weight ($n$ is carrier density, $m^*$ is carrier effective mass), and the scattering rate $1/\tau= 1/\tau _{0} +1/\tau _{i}$, representing sum of elastic ($\tau_0$, residual impurity/defect scattering) and inelastic ($\tau_i$, scattering on phonons and/or magnons) contributions.
The anisotropy in the nematic phase can be caused by both Fermi surface anisotropies arising, for instance, from the ferroorbital order triggered at the nematic transition
\cite{ferroorbital1,ferroorbital2,ferroorbital3} and/or anisotropic scattering rates, both elastic, due to the development of local magnetic order around an impurity \cite{elastic1,elastic2}, and inelastic, due to the scattering of electrons by anisotropic magnetic fluctuations \cite{inelastic1,inelastic2}  known to exist above $T_s$ \cite{neutrons}.

Stress-dependent optical reflectivity studies in Co-doped
BaFe$_2$As$_2$ (BaCo122 in the following) point to a dominant effect of the Drude
weight \cite{optics1,optics2} in the development of in-plane anisotropy. However, reconstruction of the Fermi surface by additional periodicity introduced by stripe magnetic order below the magnetic transition severely complicates the analysis \cite{folding1,folding2}.
Anisotropic reconstruction of the Fermi surface and the appearance of small pockets of high mobility carriers (similar to ``Dirac cones")
\cite{Diraccones1,Diraccones2,Diraccones3}, may dramatically alter the resistivity anisotropy \cite{anisotropyDirac,anisotropyDirac2}. Disentangling these contributions is fundamental to reveal the origin of the resistivity anisotropy in the nematic state.

Systematic studies of resistivity anisotropy over different materials provide important insight into this complicated problem.
Evolution of the resistivity anisotropy with chemical substitution was studied in the model $\mathrm{BaFe_{2}As_{2}}$ pnictide, for the cases of electron doping (Co substitution for Fe) \cite{FisherScience1}, hole doping (K substitution for Ba) \cite{Chen,BlombergNature}, as well as isoelectron substitution of As by P \cite{FisherP} and of Fe by Ru \cite{UchidaRu}. The results in both magnetically ordered state below $T_m$ and in the nematic state above show striking electron-hole asymmetry: whereas in the former the resistivity anisotropy is strongly
enhanced by increasing the Co-doping concentration, in the latter
the resistivity anisotropy is very small, and even changes sign from $\Delta \rho \equiv \rho_b-\rho_a >0$ with increasing K-doping concentration to $\Delta \rho <0$. Similar inverse anisotropy is found in other hole-doped compositions, (Ca,Na)Fe$_2$As$_2$ \cite{CaNa} and Ba(Fe,Cr)$_2$As$_2$ \cite{Crdoping}.

On the other hand in Fe(Se,Te) system, the anisotropy sign is similar to hole doped BaFe$_2$As$_2$ for both terminal compositions, FeTe \cite{FeTe} and FeSe \cite{FeSe}. The latter case is of particular interest since here tetragonal-to-orthorhombic transition is not accompanied by magnetic ordering and thus Fermi surface reconstruction. This enabled us to find that inelastic scattering provides dominant contribution to in-plane resistivity anisotropy of this material.

Study of iso-electron substitution systems provides an important advantage, since here electronic structure to zero degree remains composition independent. Unlike the Co and K substitutions, Ru
does not introduce charge carriers in the Ba(Fe$_{1-x}$Ru$_x$)$_2$As$_2$ (BaRu122 in the following). ARPES study has shown that for a wide range of Ru concentration $x$, the Fermi
surface properties -- such as the Fermi velocity and the Fermi wave-vector -- remain practically unchanged with respect to those of the parent compound \cite{Kaminski}. Thus, $\mathrm{Ba(Fe_{1-x}Ru_{x})_{2}As_{2}}$ is an ideal compound to disentangle the contributions to the resistivity anisotropy
arising from changes in the Fermi surface and from changes in the impurity concentration.

L.~Liu {\it et al.} has recently found that resistivity anisotropy of the carefully annealed samples of BaRu122 is negligible in $T=0$ limit. They suggest that the increasing anisotropy with Ru composition $x$ is due to anisotropic impurity scattering, a conclusion different from our finding of dominant contribution of inelastic scattering in FeSe \cite{FeSe}. In this study we revisit in-plane anisotropy in iso-electron substituted BaRu122 using crystals of significantly higher residual resistivity. We found that the difference between the two components of in-plane resistivity, $\rho_b-\rho_a$, obeys Matthiessen rule in a broad composition range of dirty BaRu122, but their ratio $\rho_a/\rho_b$ remains strongly temperature dependent. We interpret this observation as an indication of strong band-dependent anisotropy of the elastic scattering, negligible for high mobility carriers and sizable for low mobility carriers.

\section{Experimental}

Single crystals of Ba(Fe$_{1-x}$Ru$_x$)$_2$As$_2$ were grown using high temperature FeAs flux technique \cite{Thaler1}. The samples used in this study have relatively low Ru content and do not suffer from inhomogeneous dopant distribution. Sample compositions were determined with wavelength dispersive x-ray electron microprobe analysis (WDS) using a JOEL JXA-8200 electron-microprobe. Measurements were performed on 12 spots on sample surface. The compositions presented here are average with variance of approximately 0.2\%. Most of the samples were not heat treated after growth, we refer to these samples as as-grown in the following. For reference purpose we also used samples of pure BaFe$_2$As$_2$, annealed
at 800C for 24 hours \cite{RWHu}. These samples showed temperature dependent resistivity with residual resistivity ratio (RRR) $\sim$5,  intermediate between as-grown samples (RRR$\sim$3) and long-term annealed samples,  RRR$\sim$30, studied in Refs.~\onlinecite{anisotropyDirac2,UchidaRu}. We refer to these samples as partially annealed in the following.

White - light, optical images were taken at temperatures down to 5~K using a polarization microscope {\it Leica DMLM} with the polarizer and analyzer almost in the crossed position, as described in detail in Ref.~\onlinecite{domains}.
For initial sample screening, imaging was performed at the base temperature of $\sim$5~K, significantly below the temperatures of the coincident structural magnetic transitions $T_{sm}$  for all compositions studied. We used 50 to 100 $\mu m$ thickness slabs cleaved out of the crystals, and selected slabs with the clearest domain patterns.  Sample bars for resistivity measurements were cut using a precision wire saw with typical dimensions of 1~mm wide, 4~mm long and 0.1 mm thick, the long direction being parallel to the tetragonal [110] crystal direction, which below $T_{sm}$ becomes either the orthorhombic
$a_{o}$ or $b_{o}$ axes.

\begin{figure}[t]
\centerline{\includegraphics[width=0.4\textwidth]{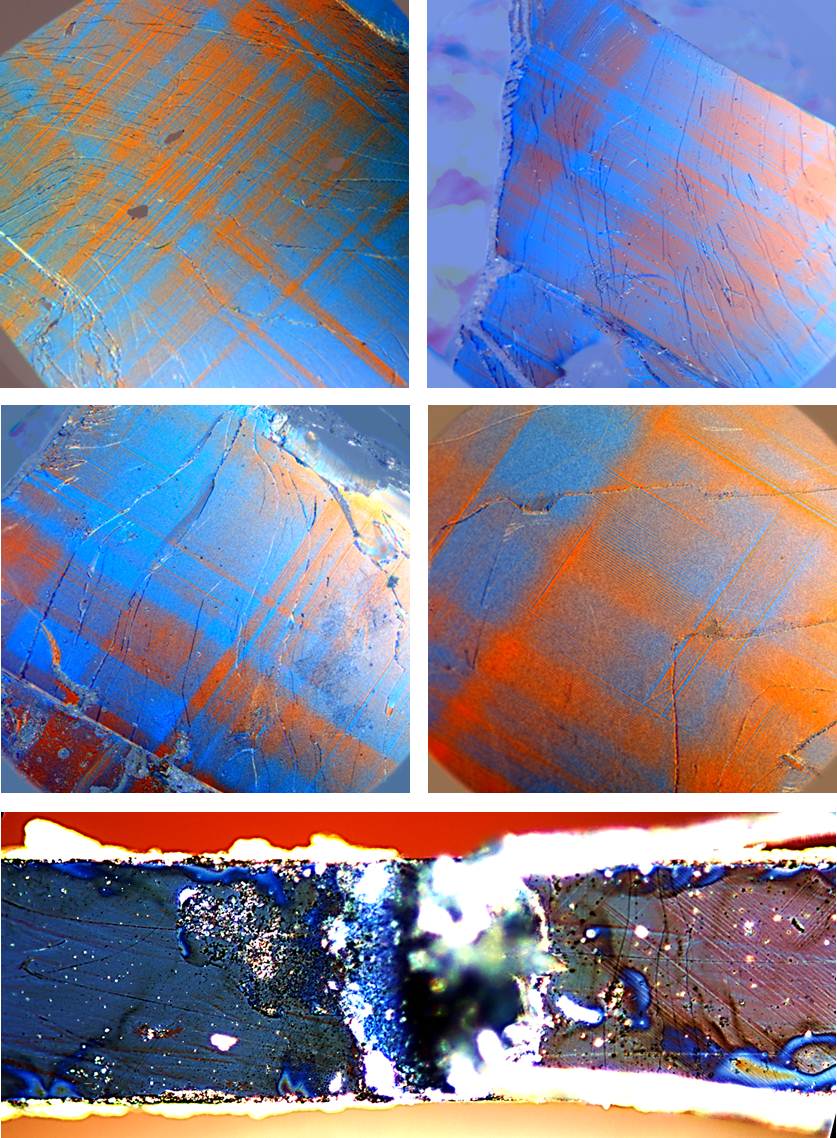}}
\caption{Top four panels: Low-temperature polarized optical microscopy images of fresh-cleaved samples of $\mathrm{Ba(Fe_{1-x}Ru_{x})_{2}As_{2}}$ with $x$ =0.048 (top left), 0.073 (top right), 0.093 (middle left) and 0.125 (middle right) reveal clear patterns of domains in free standing samples. The imaging was taken at 5~K, significantly below $T_{sm}$ for all samples.
Bottom panel: An image of a part of strain-detwinned sample. The strain is applied though potential contacts near the middle of the image. The region
to the right of the contact is under no strain (twinned) and domain
walls appear as stripes running at roughly 45 $^{o}$ to the panel.
The region left of the contact is under sufficient strain to detwin
the sample and as such no domain walls are visible. The straining
contacts also act as voltage leads for a four probe resistivty measurement
and consequently the resistivity is measured in the strain-detwinned
region only.}

\label{imaging}
\end{figure}

Contacts to the samples were made by soldering 50 $\mu m$ Ag wires using Sn \cite{SUST,patent}. Strain was applied using push-screw horse-shoe device \cite{detwinning} through potential wires, as described in detail in our previous papers \cite{detwinning,detwinningSr}. Samples remained under strain through the whole thermal cycle of resistivity measurements and during optical imaging, used to verify the completeness of the detwinning.
In the bottom panel of Fig.~\ref{imaging} we show images of domains in the strain-free portion of the samples between current and potential contacts, as well as domain-free area between potential contacts. Four-probe resistivity measurements were made in {\it Quantum design} PPMS.

The detwinning process was performed as follows: Temperature dependent
resistivity measurements and polarized optical microscopy images,
at 5 K, were taken first on unstrained samples. Strain was then applied
to samples in small, increasing increments. After each increase, samples were imaged under polarized microscopy and temperature dependent
resistivity measurements were made. As the strain increases, the domain
orientation with the longer, orthorhombic $a_{o}$-axis, parallel
to the strain becomes energetically more favorable than the other
three domain orientations and consequently occupies an increasing
fraction of the sample volume with increasing strain.

This process was continued until domains were no longer visible under
polarized microscopy and resistivity anisotropy at low temperatures saturated (very similar to that found in previous study, Fig.~1 of Ref.~\cite{UchidaRu}). Our previous X-ray diffraction studies \cite{detwinning,detwinningSr} have
shown that when sufficient strain has been applied to a sample that
domains are no longer visible, more than 90\% of the sample volume
fraction is represented by the dominant domain and is called the \textit{detwinned}
state.

In the twinned (unstrained) state, the sample is comprised of an equal
population of each domain orientation and consequently the resistivity
is the average resistivities along the $a_{o}$ and $b_{o}$-axes,
which we denote as $\rho_{t}$. In the detwinned (strained) state,
the sample is almost exclusively comprised of domains whose orthorhombic
$a_{o}$-axis is aligned parallel to the strain and therefore the
direction of current flow for the resistivity measurement, which we
denote as $\rho_{a}$. From this we may calculate the resistivity
along the orthorhombic $b_{o}$-axis as $\rho_{b}=2\rho_{t}-\rho_{a}$.

\section{Results and Discussion}

\subsection{Temperature-dependent resistivity}

\begin{figure}[h]
\centerline{\includegraphics[width=0.45\textwidth]{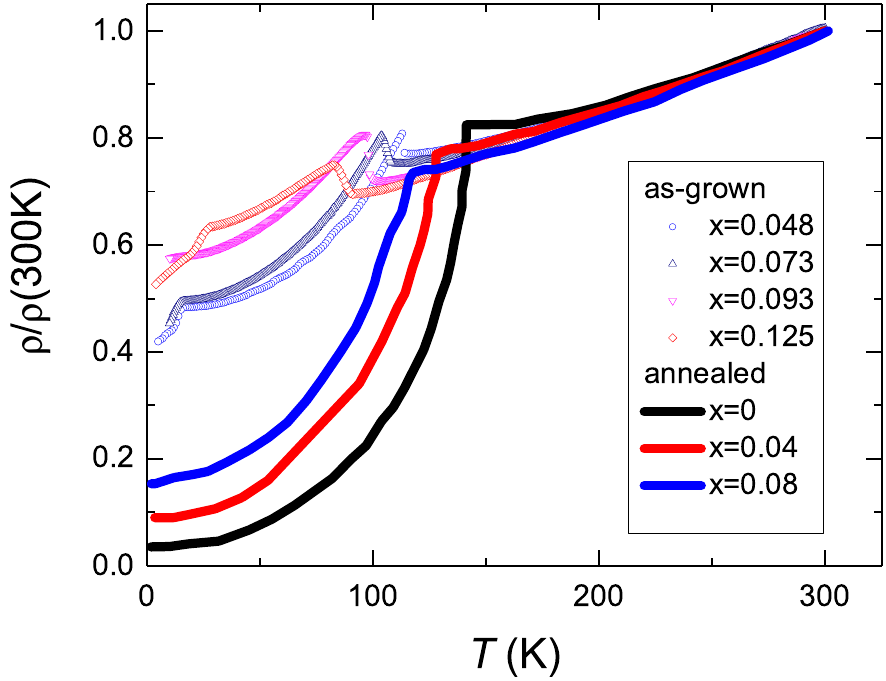}}
\caption{(Color online) Temperature-dependent resistivity, shown on normalized $\rho(T)/\rho(300K)$ scale, of the free-standing (twinned) as-grown samples of Ba(Fe$_{1-x}$Ru$_x$)$_2$As$_2$, $x$=0.048 (blue circles), $x$=0.073 (black up-triangles), $x$=0.093 (magenta down-triangles) and $x$=0.125 (red diamonds). For reference we show data for annealed samples from Ref.~\onlinecite{UchidaRu} for samples with $x$=0 (black line), $x$=0.04 (red line) and 0.08 (blue line). A significant difference in the residual resistivity (in $T=0$ limit) of the annealed and as-grown samples is found despite very close slopes of the curves above 250~K and close actual resistivity values at $T=$300~K. }

\label{normalizedresistivity}
\end{figure}

In Fig.~\ref{normalizedresistivity} we show temperature-dependent resistivity of the as-grown samples of BaRu122 used in this study, with $x$=0.048, 0.073, 0.093 and 0.125. To facilitate the comparison, we plot the data using normalized resistivity scale, $\rho /\rho(300K)$, which removes error of geometric factor determination resulting in approximately 10\% variation of $\rho(300K)$. Most of the samples  show filamentary superconductivity below 20~K, related to inhomogeneous strain distribution \cite{Paglione,Colombier,SKKim}. For reference we plot similar measurements by L.~Liu {\it et al.} on annealed samples \cite{UchidaRu}. Two differences between as-grown and annealed samples should be pointed. As naturally expected, as-grown samples have significantly higher residual resistivity and hence lower residual resistivity ratio, RRR=$\rho(300K)/\rho(0)$. However, this difference is clearly visible only in the data below the temperature of the structural/magnetic transition $T_{sm}$, seen as a clear anomaly in $\rho(T)$. The temperature of the structural/magnetic transition itself in as-grown samples is always lower than in the annealed samples,
and $T_{sm}$ is suppressed more rapidly by combined action of substitution/disorder than by Ru substitution itself. This is very similar to the known effect of artificial disorder introduced by electron irradiation \cite{Kyuil,TcTN} on the temperatures of structural and superconducting transitions.

Significant variation of the residual resistivity of the samples, however, does not affect much resistivity of the samples at room temperature, in gross violation of Matthiessen rule. The actual value of resistivity $\rho(300K)$ remains unchanged within error bars of our measurements (mainly determined by the statistical error of geometric factor determination in prone to cracks micaceous samples of BaFe$_2$As$_2$ based materials \cite{anisotropy,pseudogap}) for all Ru concentrations \cite{Rucaxis}, and coincides within the same uncertainty with the value determined by L.~Liu {\it et al.} for annealed samples \cite{UchidaRu}. Moreover, normalizing slopes of the $\rho(T)$ curves in the interval 250 to 300~K (the range where no strain induced anisotropy is observed) we obtain same, within 1\% accuracy of the procedure, offset of the curves. For all the following analysis we accepted $\rho(300K)$=300 $\mu \Omega$cm for samples of all $x$.

Observation of gross Matthiessen rule violation is in line with expectation for conductivity mechanism change in the ordered state below $T_{sm}$ due to electronic structure modification. It reflects both change (decrease) of the carrier density with respect to paramagnetic state above $T_{sm}$ and appearance of novel type of high mobility carriers, Dirac fermions, formed in the nodal areas of spin-density wave gap \cite{nodalSDW}. Detailed Shubnikov-de Haas oscillation measurements on detwinned crystals of the parent BaFe$_2$As$_2$ in the magnetically ordered state reveled one hole and two electron pockets of the Fermi surface \cite{folding1}, with all pockets being three-dimensional in nature. The electrons belonging to $\gamma$ pocket are located close to band-crossing points referred to as Dirac points. Due to high mobility these carriers are dominating the conductivity of annealed samples at low temperatures \cite{Diraccones1, Diraccones2, Diraccones3,anisotropyDirac,anisotropyDirac2}, and are responsible for huge high-field magnetoresistance \cite{Diraccones1,Diraccones2}, strongly diminished in disordered samples \cite{folding1}.

For our following discussion it is important to understand that disorder selectively suppresses contribution of high mobility carriers. The conductivity of a metal with multiple sheets of the Fermi surfaces is determined by a sum of partial
conductivities,
$\sigma =\Sigma \sigma _i= \Sigma e(n_i/m_i^*)\tau_i$.
For simplicity lets consider a metal with two types of carriers. Carriers of type 1 are usual carriers, with density $n_1$, Fermi velocity $v_1$, effective mass $m_1^*$ and scattering time $\tau_1$. Carriers of type 2 represent Dirac fermions, and are characterized by $n_2 \ll n_1$, $m_2^* \ll m_1^*$, $v_2 \gg v_1$. As was shown by Kuo {\it et al.} \cite{anisotropyDirac} from the analysis of the normal state magnetoresistance, the two contributions are of similar magnitude in non-annealed samples. Despite $n_1 \ll n_2$, conductivities $\sigma_1$ and $\sigma_2$ may be comparable due to a difference in effective masses/Fermi velocities. For samples with impurities, however, scattering rates for impurity, $1/\tau_{0}$ and inelastic $1/\tau_i$, processes are added as $1/\tau=1/\tau_{0}+1/\tau_{i}$, with $\tau$ saturation at $\tau_{0}$ in residual resistivity range.  In samples with disorder, scattering rates on impurities for each type of carrier are dramatically different. The mean free path of the carriers in residual resistivity range is of the order of inter-impurity distance, $l_0$, and thus $\tau _0=l_0/v$. Due to $v_2 \gg v_1$, this leads to $\tau_{0,1} \gg \tau_{0,2}$. As a result, in relatively dirty samples scattering rate on impurities for high mobility carriers becomes significantly higher than inelastic scattering rate. It is natural to assume that  inelastic scattering rates for two types of carriers $\tau_{i,1}$ and $\tau_{i,2}$ are similar, though it is not necessary the case for strongly $Q$-dependent magnetic scattering. That is why once $\tau_0$ determines conductivity of high mobility carriers $\sigma_2 \ll \sigma_1$ and is effectively diminished.

Important conclusion of this discussion is that even in a metallic state, without change of the band structure, the Matthiessen rule can be obeyed only as long as one carrier type is dominating the total conductivity. It can be obeyed approximately if the two types of carriers have similar properties and thus similar $\tau_0$. Thus analyzing temperature-dependent resistivity as a function of disorder, we can highlight different contributions to conductivity and scattering. Detwinned samples give an additional bonus in this respect, here the difference between $\rho_a$ and $\rho_b$ would to a notable extent diminish contribution of inelastic scattering events.

\subsection{Anisotropic resistivity}

\begin{figure}[h]
\centerline{\includegraphics[width=0.45\textwidth]{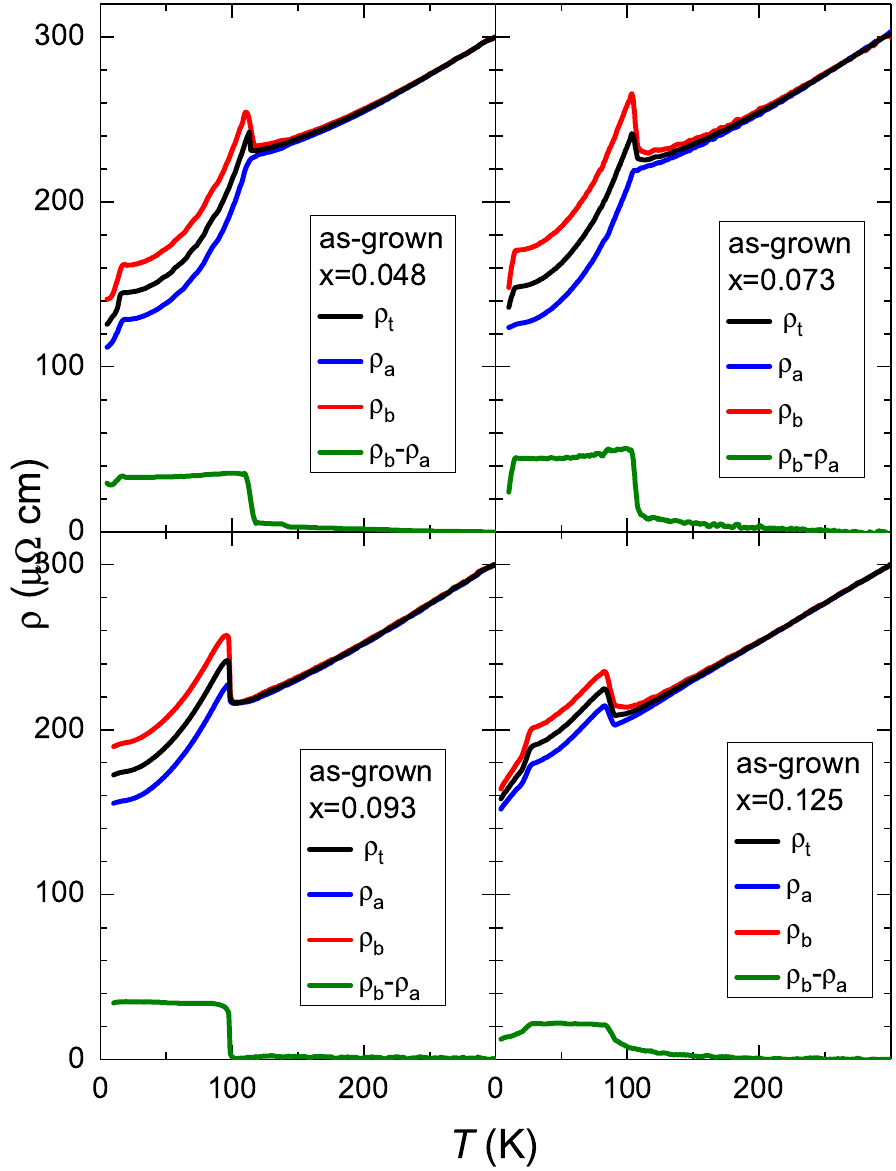}}
\caption{(Color online) Temperature-dependent resistivity of the as-grown samples of Ba(Fe$_{1-x}$Ru$_x$)$_2$As$_2$ in the free-standing (twinned) state (black curves), in strain-detwinned state, $\rho_a(T)$ blue curves, and calculated resistivity along short orthorhombic direction, $\rho_b(T)$ red curves, and a difference, $\Delta \rho (T) \equiv \rho_b(T)-\rho_a(T)$ green curves, for samples with  $x$=0.048 (top left panel), $x$=0.073 (top right panel), $x$=0.093 (bottom left panel) and $x$=0.125 (bottom right panel). }

\label{detwinnedRudoped}
\end{figure}

\begin{figure}[h]
\centerline{\includegraphics[width=0.45\textwidth]{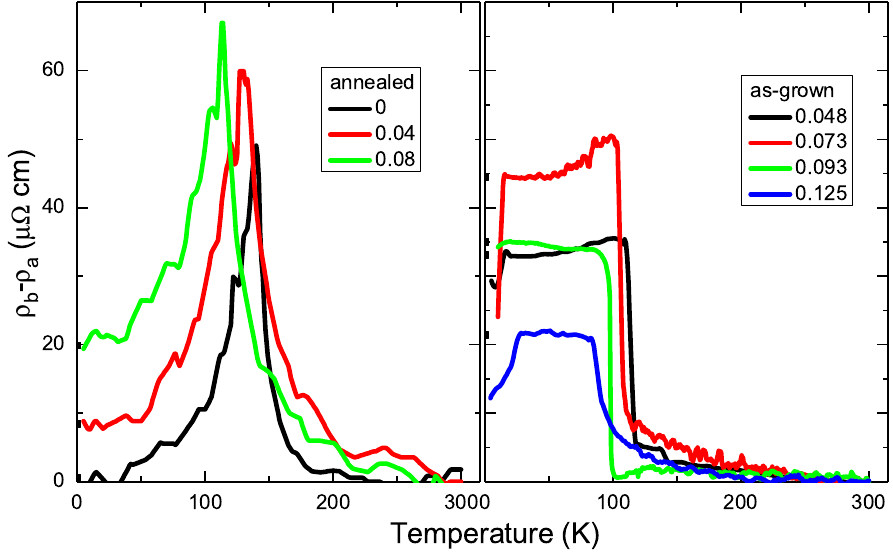}}
\caption{(Color online) Temperature-dependent difference of in-plane resistivities, $\Delta \rho (T)\equiv \rho_a(T)-\rho_b(T)$, for detwinned annealed samples of Ba(Fe$_{1-x}$Ru$_x$)$_2$As$_2$, data from Ref.~\onlinecite{UchidaRu} (left panel, $x$=0 black curve, $x$=0.04 red curve, and $x$=0.08 green curve) and for as-grown samples (right panel, $x$=0.048 black line, $x$=0.073 red line, $x$=0.093 green line, and $x=$0.125 blue line). Note practically temperature-independent $\Delta \rho (T)$ below structural/magnetic transition temperature in as-grown samples $x$=0.093 and $x$=0.125 with highest residual resistivity.
}

\label{difference}
\end{figure}

\begin{figure}[h]
\centerline{\includegraphics[width=0.45\textwidth]{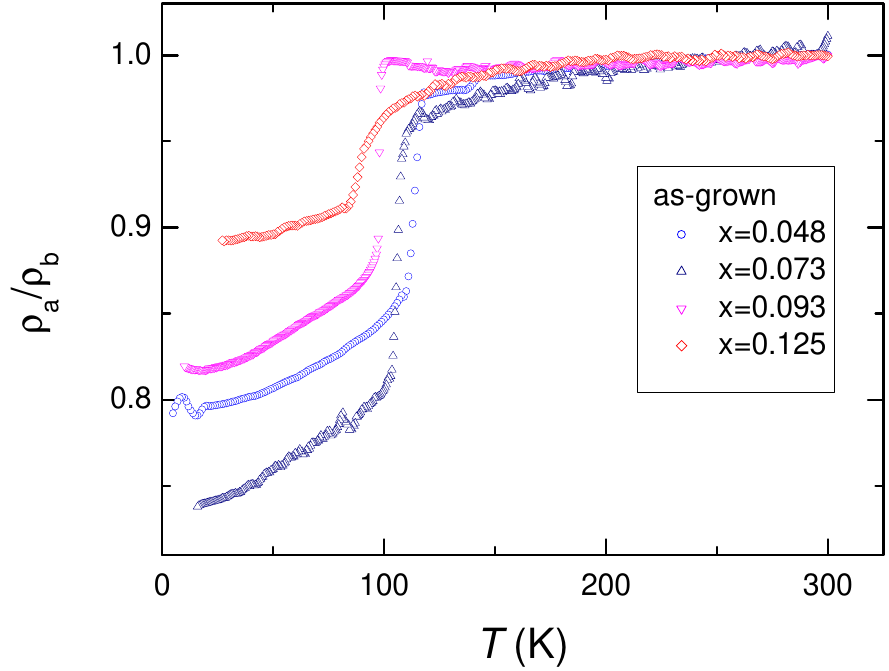}}
\caption{(Color online) Temperature-dependent ratio of in-plane resistivities, $\rho_a(T)/\rho_b(T)$, for as-grown samples of Ba(Fe$_{1-x}$Ru$_x$)$_2$As$_2$ with $x$=0.048 black line, $x$=0.073 (red line, $x$=0.093 (green line, and $x=$0.125 (blue line). The ratio shows sharp increase on cooling below structural/magnetic transition temperature $T_{sm}$ reflecting change of the carrier density and mild dependence on cooling to $T \to $0.
}

\label{ratio}
\end{figure}

In Fig.~\ref{detwinnedRudoped} we plot temperature-dependent anisotropic resistivity as measured in strain-free twinned state, $\rho_t(T)$, strained state, $\rho_a(T)$, and calculated $\rho_b(T)$ and the difference $\Delta \rho(T)=\rho_b-\rho_a$. The data are shown for samples with $x$=0.048 (top left panel), $x$=0.073 (top right), $x$=0.093 (bottom left) and $x$=0.125 (bottom right). Of note that the difference $\Delta\rho(T)$ remains almost temperature independent in as-grown samples of all compositions.

In Fig.~\ref{difference} we plot temperature-dependent resistivity difference $\Delta \rho \equiv \rho_b-\rho_a$ for annealed samples of BaRu122, $x$=0, 0.04 and 0.08, data from Ref.~\onlinecite{UchidaRu} and as-grown samples with $x$ up to 0.125. The difference between the two data sets is dramatic. In disordered samples the difference remains nearly temperature-independent, while in annealed pure samples $\Delta \rho $ takes a sharp maximum below $T_{sm}$ and decreases to zero on further cooling. Addition of Ru leads to build-up of finite residual difference at $T=$0 in annealed samples. These results clearly suggest that high mobility carriers which dominate conductivity of pure samples at low temperatures, do not have any anisotropy of the residual resistivity, while ``normal" carriers do. The temperature dependent part of resistivity in pure samples comes from ``normal" carriers close to $T_{sm}$ and from high mobility carriers at $T$ going to zero. Note also that the difference immediately below $T_{sm}$  is comparable in annealed and as grown samples.

As we discussed above, observation of essentially temperature-independent difference for two components of in-plane resistivity is strong indication that one band is dominating transport of dirty samples, and conductivity of high mobility carriers is essentially eliminated. In situation of one dominant band we can distinguish between anisotropy of transport determined by Drude weight (Fermi surface) and by scattering.
The Drude weight enters as a multiplicative factor into the expression for conductivity, and thus we should expect temperature independent $\rho_a/\rho_b$ ratio. In Fig.~\ref{ratio} we plot ratio of the component resistivities in disordered samples of BaRu122. In all cases the ratio shows rapid decrease from $\rho_a/\rho_b$=1 above $T_{sm}$ to lower values immediately below transition, followed by monotonic decrease on cooling. Observation of temperature dependent ratio of the in-plane resistivity components in conjunction with temperature -independent difference, see right panel of Fig.~\ref{difference}, strongly suggests that the anisotropy is determined by residual resistivity.

\begin{figure}[h]
\centerline{\includegraphics[width=0.45\textwidth]{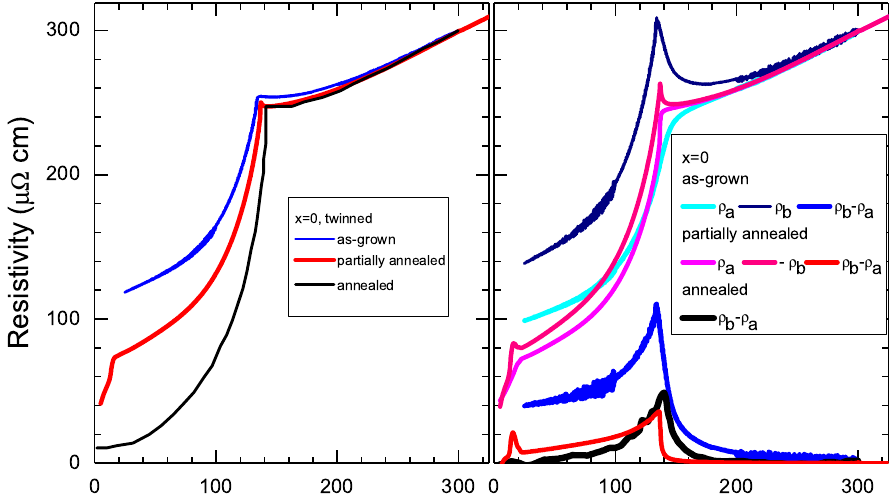}}
\caption{(Color online) Left panel: Temperature-dependent resistivity of free standing (twinned) samples of pure BaFe$_2$As$_2$ in as-grown (blue curve, data from Ref.~\cite{detwinning}), partially annealed (red curve) and strongly annealed (black curve, data from Ref.~\onlinecite{UchidaRu}). Right panel shows anisotropic in-plane resistivity measured for detwinned samples of as-grown (cyan curve for $\rho_a$ and navy curve for $\rho_b(T)$) and partially annealed samples (magenta curve for $\rho_a$ and pink curve for $\rho_b(T)$) of pure BaFe$_2$As$_2$ and the difference of in-plane resistivities, $\Delta \rho (T)\equiv \rho_a(T)-\rho_b(T)$, for as-grown (blue curve), partially annealed (red curve) and annealed (black curve) samples. Strong anisotropy of the resistivity in as-grown samples above $T_{sm}$ is indicative of excessive stress in early study \cite{detwinning} .
}
\label{pureannealing}
\end{figure}

To get an additional insight into suggested switchover, in Fig.~\ref{pureannealing} we compare temperature-dependent anisotropic resistivity of pure BaFe$_2$As$_2$ with different annealing conditions. We plot data for as grown sample, intermediately annealed sample and carefully annealed sample (data from Ref.~\onlinecite{UchidaRu}). Left panel shows resistivity of the samples in the twinned state, revealing monotonic decrease of residual resistivity with improved annealing. Right panel shows data in the detwinned state as well as the temperature-dependent difference. As can be seen, the residual term in the difference $\Delta \rho(0)$ rapidly increases with disorder, confirming the trend found in Ru-doped samples. Of interest, the magnitude of $\Delta \rho$ increase below $T_{sm}$ does not show monotonic dependence. The difference $\Delta \rho (T_{sm})$ is the largest in the samples with the biggest effect above $T_{sm}$. This observation may suggest that the difference may come from a difference in the detwinning conditions, i.e. the value of compressive, as in Ref.~\onlinecite{UchidaRu}, or expansive stress used to detwin the samples.

\begin{figure}[h]
\centerline{\includegraphics[width=0.45\textwidth]{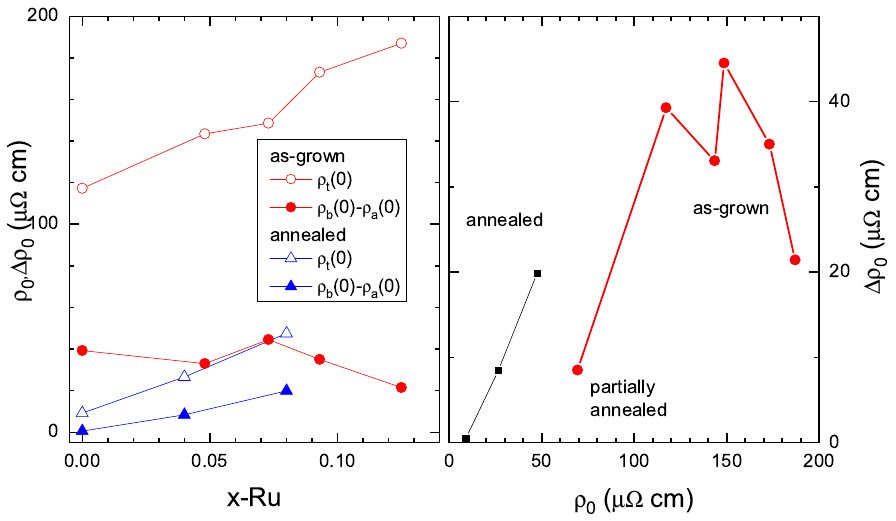}}
\caption{(Color online) Left panel: Low temperature $T \to $0 extrapolation of residual resistivity, $\rho (0)$, open symbols, and residual resistivity difference, $\Delta \rho(0)$ closed symbols, shown as a function of $x$ in samples of  Ba(Fe$_{1-x}$Ru$_x$)$_2$As$_2$ in as-grown (red curves) and annealed (blue curves, Ref.~\onlinecite{UchidaRu}) states. Right panel shows $\Delta \rho(0)$  as a function of $\rho(0)$, revealing non-monotonic dependence of the two. Rapid rise in the curve is followed by a general trend to decrease with some non-monotonic dependence in the same $x$ range where non-monotonic trend in thermopower evolution with $x$ was reported \cite{Halyna}.
}
\label{summary}
\end{figure}

In Fig.~\ref{summary} we summarize our observations plotting $\Delta \rho (0)$ and $\rho (0)$ as a function of $x$ (left panel) for as-grown, partially annealed and annealed samples. We also plot $\Delta \rho (0)$ vs $\rho (0)$, explicitly checking the linear relation between the two. In our analysis we ignored features due to partial superconductivity, observed in many samples below approximately 20~K. Two trends can be noticed. In addition to rising $\Delta \rho$ in ultra-pure samples, reflecting suppression of the contribution of high mobility carriers, a decrease of $\Delta \rho(0)$ vs $\rho(0)$ starts in heavily disordered samples.
A non-monotonic feature observed for sample $x$=0.073 coincides in $x$ with doping anomaly of Seebeck effect evolution, suggested as either originating from Fermi surface topology change (Lifshits transition) or gross change in transport scheme \cite{Halyna}. From our data we can conclude that at least this crossover is not due to suppression of Dirac fermion contribution, which happens at lower $\rho(0)$.

The data of Fig.~\ref{summary} clearly show that there are several processes involved in the evolution of resistivity anisotropy. High mobility carriers show no in-plane anisotropy of the electronic transport on $T \to 0$. This conclusion is similar to the experimental observation of negligible resistivity anisotropy in residual range in FeSe \cite{FeSe}. In FeSe all sheets of the Fermi surface are small \cite{TerashimafermiologyFeSe} and of relatively high mobility. Observation of temperature independent difference $\Delta \rho (T)$ for all dirty samples of BaRu122 as opposed to temperature dependent ratio, Figs.~\ref{difference} and \ref{ratio} implies that residual resistivity rather than the electronic structure is responsible for the anisotropy of the ``normal" carriers.

\section{Conclusion}

The main conclusions of this study may be summarized as follows. Resistivity of ultrapure annealed samples at low temperatures is mainly determined by high mobility carriers. This contribution does not show any anisotropy in residual resistivity range. Anisotropic contribution of normal mobility electrons can be clearly seen in samples with high residual resistivity, in which contribution of high mobility carriers is quenched by disorder, and in a temperature range immediately below $T_{sm}$. This contribution reveals temperature-independent difference of anisotropic resistivity components, which suggests its relation to anisotropy of residual resistivity. It does not show temperature-independent ratio of the components, as would be expected for anisotropy determined by the Fermi surface parameters. The reduced anisotropy in both elastic and inelastic channels in annealed samples is determined by the dominance in the conductivity of the high mobility carriers, strongly diminished in disordered as-grown samples.
Our finding suggests that magnetic/nematic order has dramatically different effect on different parts of the Fermi surface. It predominantly affects inelastic scattering for small pocket high mobility carriers and elastic impurity scattering for larger sheets of the Fermi surface.

\begin{acknowledgements}
The authors acknowledge useful discussions with R.~M.~Fernandes. This work was supported by the U.S. Department of Energy (DOE), Office of Science, Basic Energy Sciences, Materials Science and Engineering Division. The research was performed at the Ames Laboratory, which is operated for the U.S. DOE by Iowa State University under contract DE-AC02-07CH11358.
\end{acknowledgements}


\end{document}